\documentclass[twocolumn,prd,superscriptaddress,showpacs,floatfix,%
preprintnumbers,nofootinbib]{revtex4}

\usepackage{epsfig}

\begin{document}

\title{Are cosmological neutrinos free--streaming?}

\author{Anders Basb{\o}ll}
\affiliation{Department of Physics and Astronomy, University of
Aarhus, Ny Munkegade, DK-8000 Aarhus C, Denmark}

\author{Ole Eggers Bj{\ae}lde}
\affiliation{Department of Physics and Astronomy, University of
Aarhus, Ny Munkegade, DK-8000 Aarhus C, Denmark}

\author{Steen Hannestad}
\affiliation{Department of Physics and Astronomy, University of
Aarhus, Ny Munkegade, DK-8000 Aarhus C, Denmark}

\author{Georg G.~Raffelt}
\affiliation{Max-Planck-Institut f\"ur Physik
(Werner-Heisenberg-Institut), F\"ohringer Ring 6, 80805 M\"unchen,
Germany}

\date{9 June 2008}
\long\def\symbolfootnote[#1]#2{\begingroup%
\def\thefootnote{\fnsymbol{footnote}}\footnote[#1]{#2}\endgroup}
\preprint{MPP-2008-55}

\begin{abstract}
Precision data from cosmology suggest neutrinos stream freely and
hence interact very weakly around the epoch of recombination. We
study this issue in a simple framework where neutrinos recouple
instantaneously and stop streaming freely at a redshift $z_i$. The
latest cosmological data imply $z_i\alt 1500$, the exact constraint
depending somewhat on the assumed prior on $z_i$. This bound
translates into a limit on the coupling strength between neutrinos
and majoron-like particles $\phi$, implying $\tau \agt 1 \times
10^{10} \, {\rm s} \,(m_2/50\,{\rm meV})^3$ for the decay $\nu_2 \to
\nu_1+\phi$.
\end{abstract}

\pacs{98.80.-k, 14.60.St, 14.80.Mz}

\maketitle

\section{Introduction}                        \label{sec:introduction}

With the advent of high-precision cosmology it has become feasible to
probe progressively more detailed aspects of the cosmic neutrino
background radiation~\cite{Hannestad:2006zg, Lesgourgues:2006nd}. In
the standard model, neutrinos provide relativistic energy density
which influences the cosmic microwave background (CMB) radiation
mainly via the early Integrated Sachs--Wolfe (ISW) effect and the
matter fluctuation spectrum via the relation between neutrino energy
density and the epoch of matter--radiation equality. The existence of
a cosmological background of relativistic energy density has been
unambiguously detected in the \hbox{WMAP-5}
data~\cite{Komatsu:2008hk} and was already previously detected using
the combination of CMB and Large Scale Structure (LSS)
data~\cite{Hannestad:2001hn, Hannestad:2005jj, Ichikawa:2006vm,
Mangano:2006ur, Hamann:2007pi}. Furthermore, cosmological
data provide a restrictive upper bound on the sum of neutrino masses of
0.2--1~eV, depending on the specific choice of data sets
and model space~\cite{Zunckel:2006mt, Cirelli:2006kt, Goobar:2006xz,
Kristiansen:2006xu, Seljak:2006bg, Fogli:2006yq, Hannestad:2003xv,
Hannestad:2006zg, Hannestad:2007tu}.

The present level of precision allows us to turn to more subtle
issues. For example, it is timely to probe the possibility that
neutrinos have non-standard interactions where one case in point is an
interaction with the Nambu-Goldstone boson of a new, broken $U(1)$
symmetry as in majoron models~\cite{Gelmini:1980re,
Chikashige:1980ui, Schechter:1981cv}. Such an interaction would
recouple the neutrinos to each other at some ``interaction redshift''
$z_i$, whereas at earlier epochs they would behave in the same way as
standard-model neutrinos. For the cases of interest, this recoupling
occurs much later than the electroweak decoupling. Therefore, in the
limit of relativistic neutrinos the total energy density in the
combined fluid of neutrinos and majorons is conserved, preventing any
direct impact on cosmological observables.

However, neutrinos lose their free-streaming property if the
interaction is sufficiently strong. As a consequence, any anisotropic
stress components in the Boltzmann hierarchy are suppressed,
effectively truncating the Boltzmann hierarchy at first order,
equivalent to the equations for a perfect fluid \cite{Raffelt:1987ah,
AtrioBarandela:1996ur, Beacom:2004yd, Hannestad:2004qu,
Hannestad:2005ex, Friedland:2007vv, Sawyer:2006ju, Bell:2005dr,
DeBernardis:2008ys}. (See Ref.~\cite{Ma:1995ey} for a detailed
description of the Boltzmann hierarchy.)

The impact of neutrino free streaming on cosmological observables
was recently studied in Ref.~\cite{DeBernardis:2008ys}. The fit
parameter was the effective viscosity $c_{\rm vis}^2$, taken to be
independent of redshift. Our study is complementary in that we
assume that $c_{\rm vis}^2$ drops instantaneously at $z_i$ from the
free-streaming value 1/3 to the perfect-fluid
value~0\symbolfootnote[2]{A paper by the same authors with a more
extensive treatment of this issue is in preparation.}. Our
conclusion that neutrinos should stream freely around the epoch of
recombination is perfectly consistent with
Ref.~\cite{DeBernardis:2008ys}. However, our approach lends itself
more directly to an interpretation in terms of a specific
interaction model where the recoupling redshift is related to a
dimensionless coupling constant~$g$. Therefore, we can translate our
limits on $z_i$ into limits on~$g$.

From flavor oscillation experiments we know that neutrinos have masses
which therefore are unavoidable cosmological fit parameters.  The
usual cosmological limits on the sum of neutrino masses imply that any
single mass eigenstate should obey $m\alt0.2$--0.3~eV so that all
neutrinos would be relativistic around the recombination
epoch. Treating them as massless is therefore a reasonable
approximation for the simple problem addressed here. On the other
hand, a strong majoron-type interaction can lead to the annihilation
of ``heavy'' neutrinos into majorons (``neutrinoless
universe''~\cite{Beacom:2004yd}). Such scenarios lead to a complicated
evolution of the neutrino-majoron fluid that we are not investigating,
although it would have a strong impact on cosmological observables. In
any event, our constraint on the free-streaming nature of the relevant
radiation at recombination does not depend on the physical nature of
the radiation.

Eventually the KATRIN experiment, unless it detects a significant
neutrino mass, will constrain the neutrino mass scale to $m\alt0.2$~eV
\cite{Drexlin:2005zt}. Such a bound would imply that neutrinos can not
have disappeared at the recombination epoch and our constraint indeed
applies to neutrinos. In this sense the anticipated KATRIN limit will
strengthen the case for translating our limit on $z_i$ into a limit on
exotic neutrino interactions.

We begin in Sec.~\ref{sec:models} with a description of our model
space, data sets, and statistical methodology. In
Sec.~\ref{sec:recoupling} we provide our bounds on $z_i$ that are
translated, in Sec.~\ref{sec:g}, into limits on a neutrino-majoron
coupling strength $g$. We conclude in Sec.~\ref{sec:discussion}.

\section{Models, data, and methodology}             \label{sec:models}

Our parameter constraints will be based on a reasonably general
10-parameter model consisting of
\begin{equation}
\Theta =
(\omega_{\rm CDM},\omega_{\rm B},H_0,n_s,\alpha_s,\tau,A_s,N_\nu,z_i),
\end{equation}
where $h=H_0/(100~{\rm km}~{\rm s}^{-1}~{\rm Mpc}^{-1})$, and the
cold dark matter and baryon contents are given by $\omega_{\rm CDM} =
\Omega_{\rm CDM} h^2$ and $\omega_{\rm B} = \Omega_{\rm B} h^2$
respectively. We assume spatial flatness, i.e.\ the dark energy
density is given by $\Omega_{\rm DE} = 1 - \Omega_{\rm CDM} -
\Omega_{\rm B}$. For the dark energy we assume a constant equation of
state parameter $w$. The primordial fluctuations are assumed to be
adiabatic and described by the scalar amplitude $A_s$, the spectral
index $n_s$, and the running $\alpha_s$. We do not consider the
presence of tensor modes or an isocurvature component. Finally, we
include the present Hubble parameter, $H_0$, and the optical depth to
reionization, $\tau$. As discussed in the previous section we assume
massless neutrinos.

In order to keep our study on the properties of the radiation as
general as possible, we will sometimes use the effective number of
neutrino flavors, $N_\nu$, as a fit parameter to express the
radiation content in the usual way. The standard value is
$N_\nu=3.046$ \cite{naples}.

Neutrino interactions are assumed to recouple instantaneously at a
redshift $z_i$. Here our standard prior is linear (i.e.\ uniform) in
$z_i$, but we will also test an alternative case where a linear prior
is used on $\log(z_i+1)$.

The priors on our model parameters are listed in
Table~\ref{tab:priors}, including the alternatives that we use in
some cases.

\begin{table}[b]
\caption{\label{tab:priors}Priors on the model parameters. Note that
  for $N_\nu$ we use the prior $0 \leq N_\nu \leq 50$ for
  Figs.~\ref{fig:nz} and~\ref{fig:matter}, and $N_\nu = 3.046$
  everywhere else.}
\begin{ruledtabular}
\begin{tabular}{lll}
Parameter & Standard (linear) prior & Alternative (log) prior \\
\hline
$\omega_{\rm CDM}$  & 0.01--0.9    &  0.01--0.9 \\
$\omega_{\rm B}$    & 0.005--0.1   &  0.005--0.1 \\
$H_0$               & 40--100      &  40--100 \\
$w$                 & $-2$--0      &  $-2$--0 \\
$n_s$               & 0.5--1.5     &  0.5--1.5 \\
$\alpha_s$          & $-0.2$--0.2  &  $-0.2$--0.2 \\
$\log[10^{10} A_s]$ & 2.5--4       &  2.5--4 \\
$\tau$              & 0--1         &  0--1 \\
$N_\nu$             & 3.046 / 0--50        & 3.046 / 0--50  \\
$z_i$               & 0--$10^4$  & --- \\
$\log(1+z_i)$       & ---           & 0--4 \\
\end{tabular}
\end{ruledtabular}
\end{table}

We use CMB data from WMAP-5 \cite{lambda, Nolta:2008ih,
  Dunkley:2008ie, Komatsu:2008hk} and measurements of the matter power
spectrum based on the Sloan Digital Sky Survey--Luminous Redshift
Galaxies (SDSS--LRG) \cite{Tegmark:2006az} and 2--degree--Field (2dF)
galaxy samples~\cite{Cole:2005sx}. In addition we include the
Supernova Type~Ia (SN--Ia) data from Ref.~\cite{Davis:2007na}, the
SDSS--LRG Baryon Acoustic Oscillation measurement (SDSS--LRG BAO) from
Ref.~\cite{Eisenstein:2005}, and the Hubble Space Telescope (HST) key
project measurement of $H_0$ \cite{Freedman:2000cf}.

Our treatment of non-linear corrections to the LSS power spectra
follows the prescription given in Ref.~\cite{Hamann:2008we}, i.e., we
include data up to $k = 0.2 \, h \, {\rm Mpc}^{-1}$ and correct for
non-linearity using the shot-noise term $P_{\rm shot}$.

In order to derive constraints on our model parameters we
have modified the publicly available CAMB code~\cite{camb} to allow
for neutrino interactions and combined it with the Markov Chain Monte
Carlo software COSMOMC~\cite{cosmomc}. Credible intervals are
calculated using Bayesian inference as implemented in the GetDist
routine of COSMOMC.

\section{Limit on Recoupling Redshift}          \label{sec:recoupling}

\begin{figure}[t]
\includegraphics[width=0.75\columnwidth]{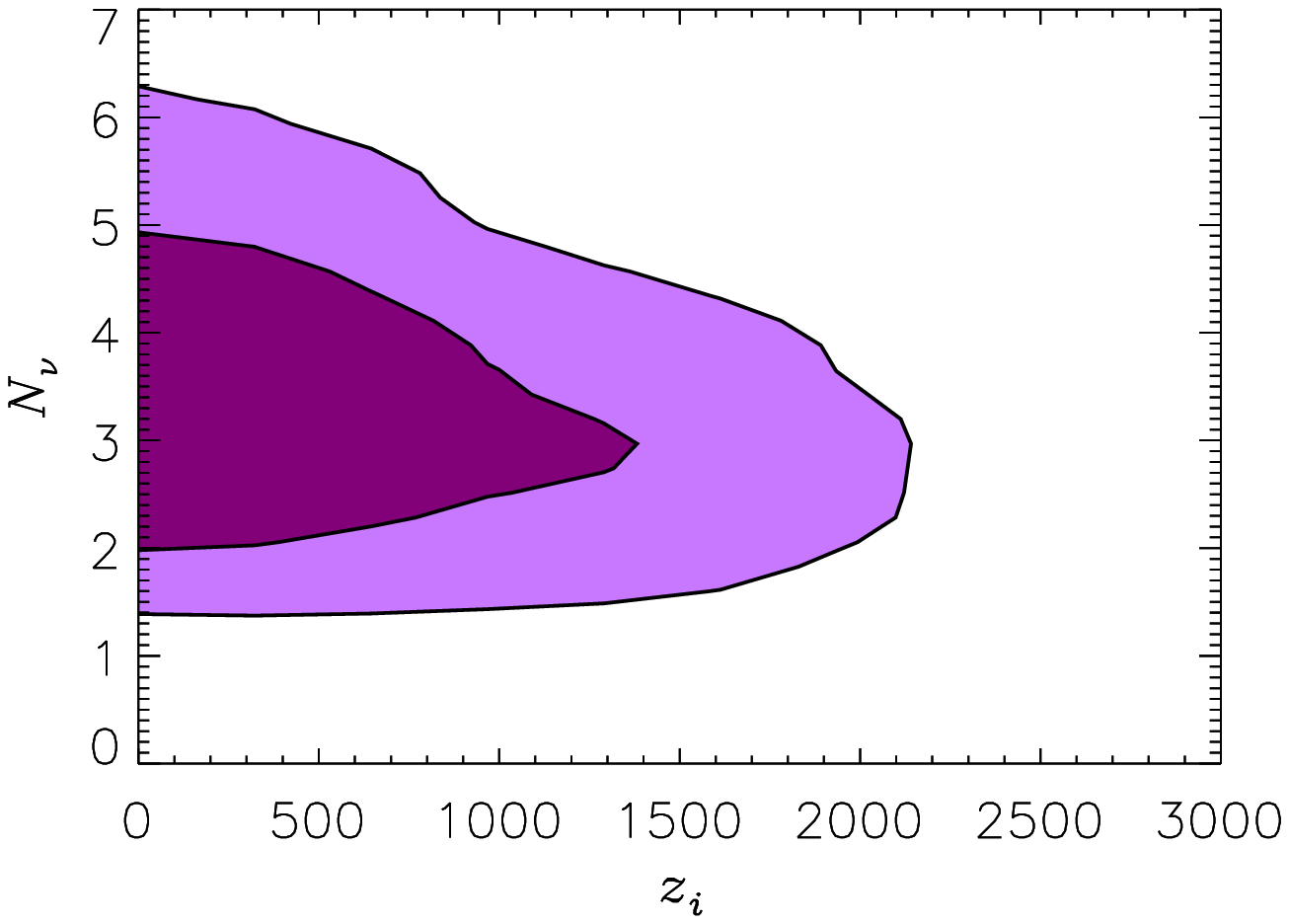}
\vskip4pt
\includegraphics[width=0.75\columnwidth]{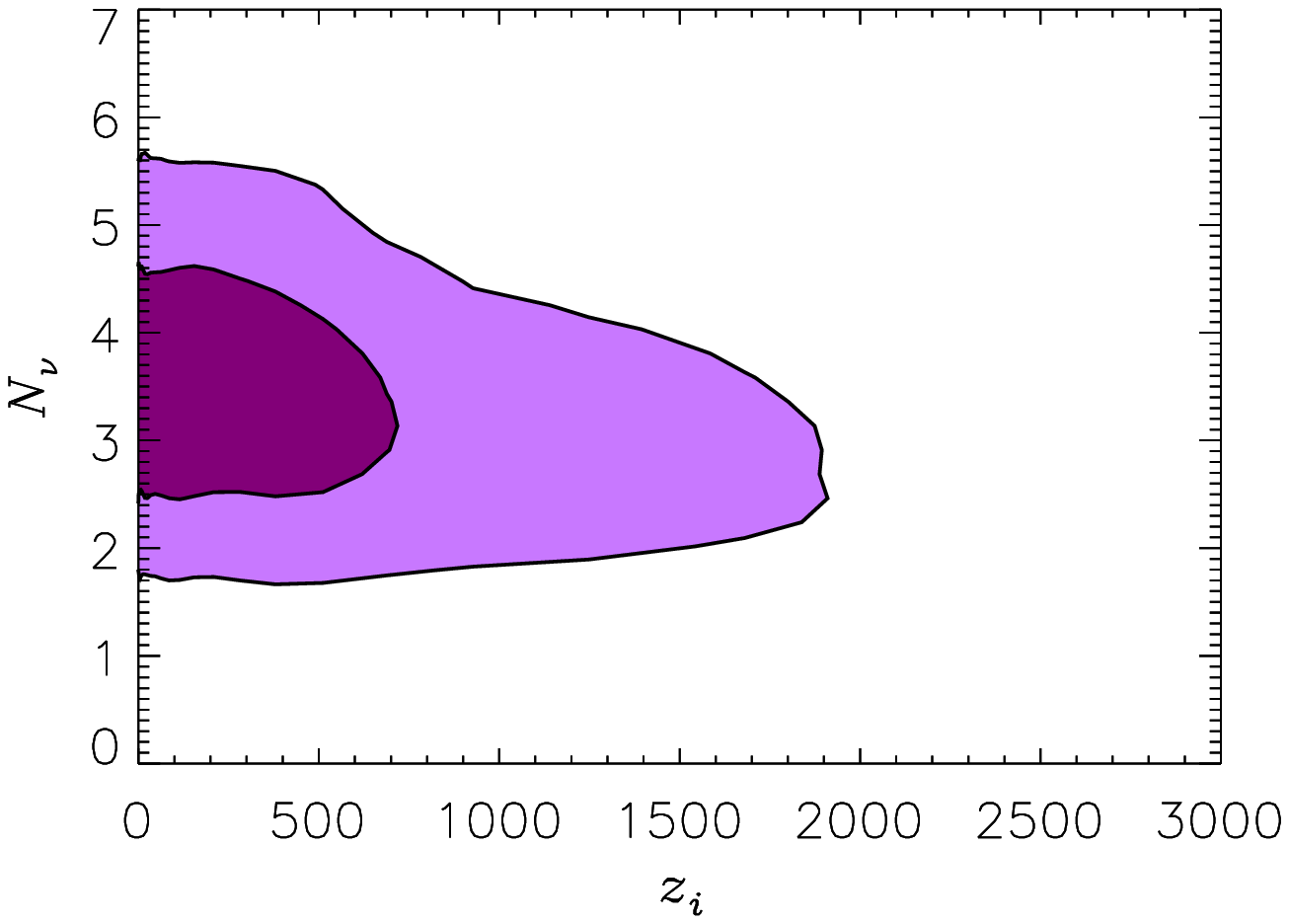}
\caption{2D marginal 68\% and 95\% contours for $z_i$ and $N_\nu$.
Top: Linear prior for $z_i$.
Bottom: Logarithmic prior, i.e., linear in
$\log(1+z_i)$.\label{fig:nz}}
\end{figure}

Following the approach described in the previous section we have
calculated 68\% and 95\% credible regions in the 2D parameter space
of $N_\nu$ and $z_i$ that we show in Fig.~\ref{fig:nz}. In the upper
panel we have used the linear prior on $z_i$ described in
Table~\ref{tab:priors}. The cosmological precision data show
(i)~strong evidence for the existence of relativistic energy density
and (ii)~that it must be freely streaming at a redshift around
recombination ($z_r\approx1100$). Marginalizing over $N_\nu$ we find
$z_i < 1500$ at 95\%~C.L.

We have repeated the same exercise for a logarithmic prior on $z_i$,
i.e., one that is uniform in $\log(z_i+1)$. The corresponding credible
regions are shown in the lower panel of Fig.~\ref{fig:nz}.
Marginalizing once more over $N_\nu$ we find $z_i < 795$ at 95\%~C.L.
The difference arises because the effective volume at low $z_i$
becomes larger for the logarithmic prior and therefore integration
favors slightly lower values of $z_i$. This effect is well known for
parameters with a highly non-Gaussian likelihood, other notable
examples being neutrino mass, $m_\nu$ \cite{Hannestad:2007tu}, and the
tensor to scalar ratio, $r$ \cite{Valkenburg:2008cz, Peiris:2006ug}.

While the exact redshift at which neutrinos can become strongly
interacting depends on assumptions about priors, we find that
neutrinos which were strongly interacting significantly before
recombination are excluded by data at much more than 95\% C.L., a
conclusion which is fully consistent with
Ref.~\cite{DeBernardis:2008ys}.

\begin{figure}[t]
\includegraphics[width=0.8\columnwidth]{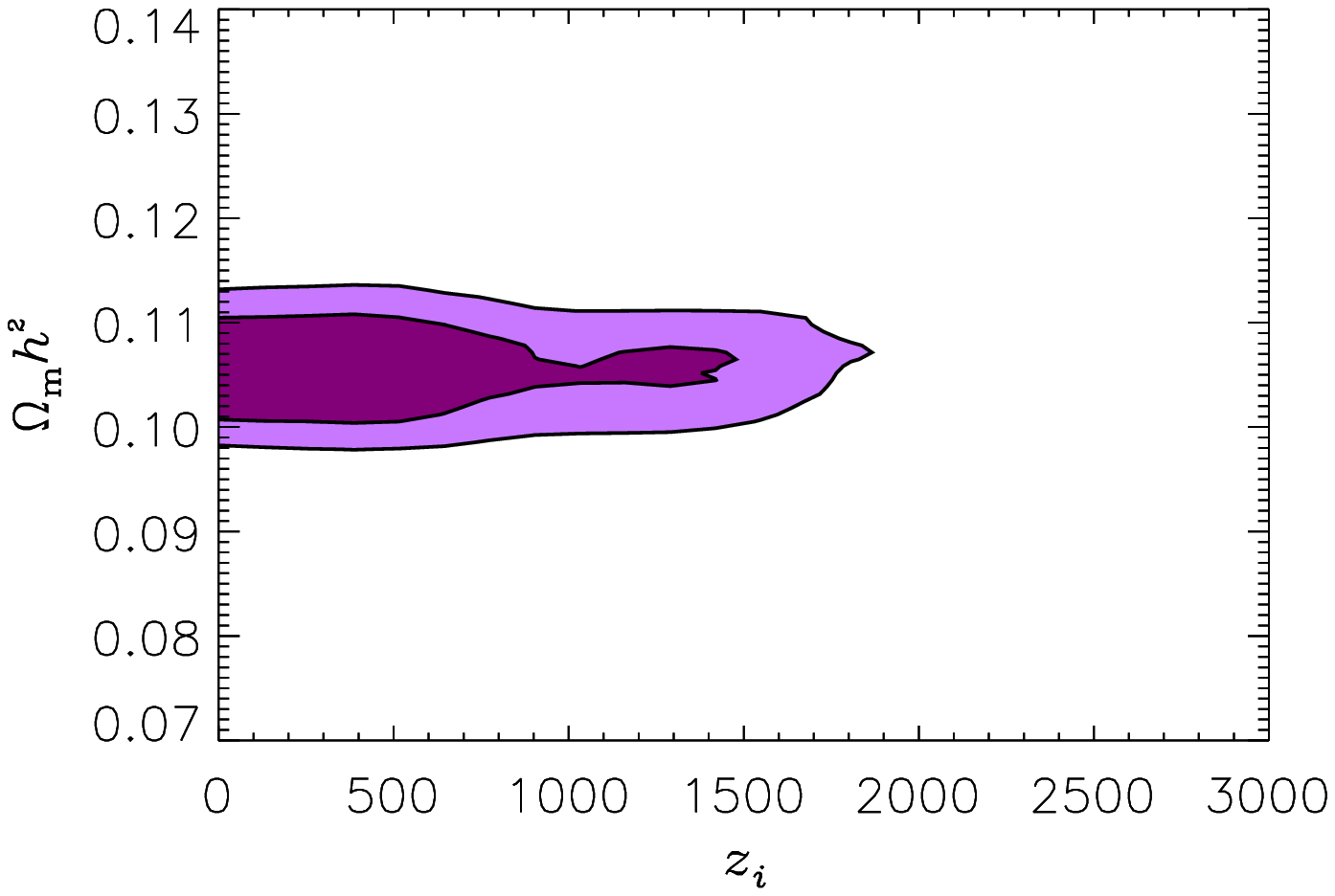}
\vskip4pt
\includegraphics[width=0.8\columnwidth]{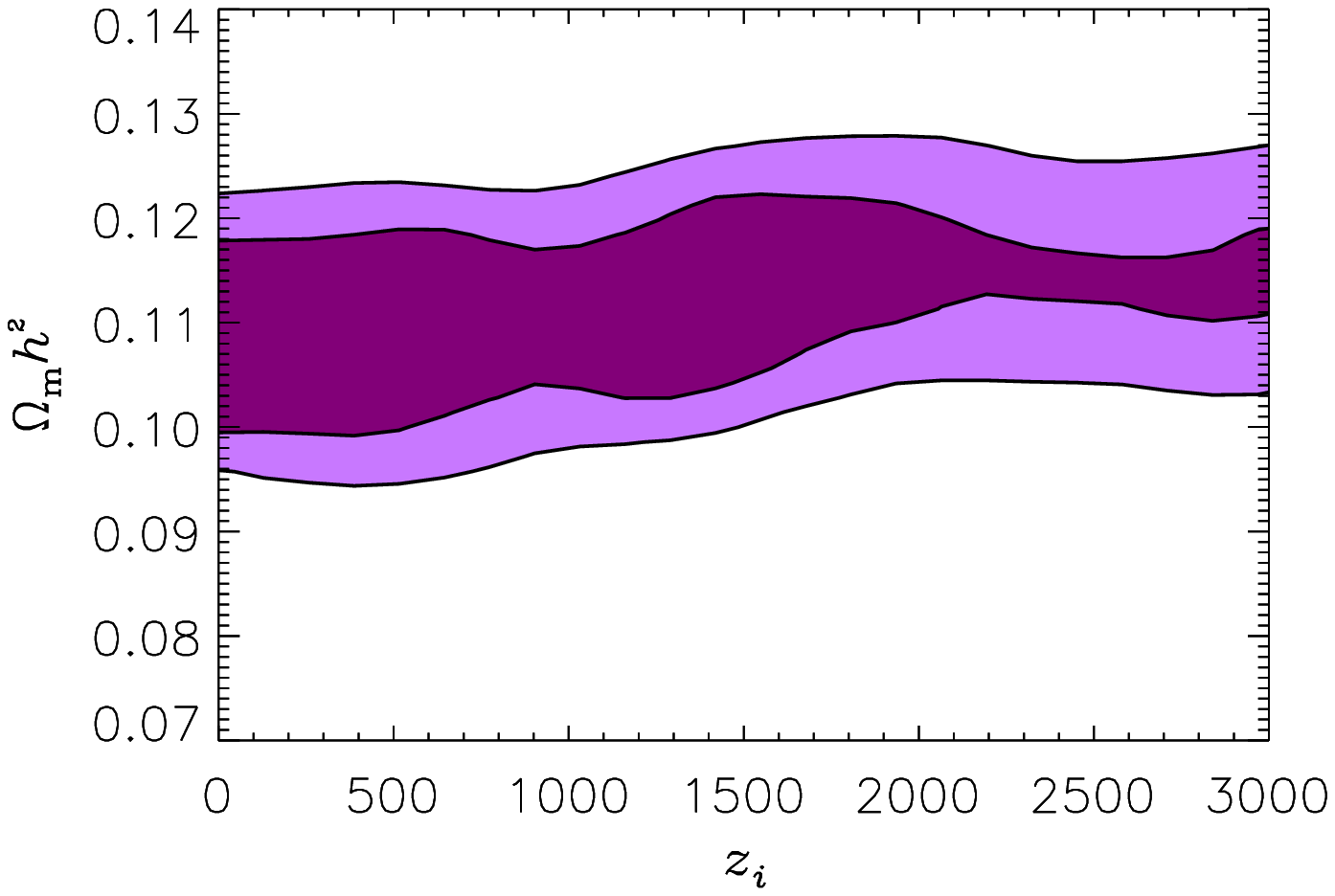}
\caption{2D marginal 68\% and 95\% contours for $z_i$ and $\Omega_{\rm
    M} h^2$, using the linear prior for $z_i$ and using $N_\nu$ as a
  fit parameter. Top: Full data set. Bottom: WMAP-5 data only.
\label{fig:matter}}
\end{figure}

Our results pertain to any form of radiation present around
recombination. However, we ultimately want to test the interactions
of ordinary neutrinos. The cosmic standard radiation content is given
by $N_\nu=3.046$. Repeating the above exercises with this fixed prior
we find $z_i < 1520$ for the linear $z_i$ prior and $z_i < 790$ for
the logarithmic prior. These limits are almost identical to those
where we marginalized over $N_\nu$. This is hardly surprising since
$N_\nu \sim 3$ allows the largest values~of~$z_i$.

In order to test more quantitatively how disfavored strongly coupled
neutrinos are we have performed a high-precision run with
$N_\nu=3.046$ and the more conservative linear $z_i$ prior to
calculate a sequence of progressively higher confidence limits. We
find $z_i < 1910$ at 99\% C.L. and $2230$ at 99.7\% C.L. At even
higher confidence limits the Markov chains show signs of incomplete
convergence and we refrain from quoting bounds.

We also show 2D credible regions in the plane spanned by the matter
density $\Omega_{\rm M}=\Omega_{\rm CDM}+\Omega_{\rm B}$ and $z_i$ in
Fig.~\ref{fig:matter} where the conservative linear $z_i$ prior was
used and $N_\nu$ kept as a fit parameter. In the top panel we have
used the full data set as in Fig.~\ref{fig:nz} and find consistent
results. In the bottom panel have used only WMAP-5 data and thus
confirm with our method that WMAP-5 data alone do not significantly
constrain~$z_i$~\cite{DeBernardis:2008ys}.
\vspace*{-0.2cm}
\section{Limit on Coupling Strength}                     \label{sec:g}
\vspace*{-0.2cm}
Our approach of neutrinos recoupling at some redshift $z_i$ was
motivated by a majoron-type interaction model where neutrinos interact
with a new massless pseudoscalar by virtue of a dimensionless Yukawa
coupling $g$. In the framework of such a model we can translate our
limit on $z_i$ into a limit on $g$ in analogy to a previous paper by
two of us~\cite{Hannestad:2005ex}. When considering the scattering
process the bound applies to any component of $g_{ij}$, the indices
referring to the different neutrino flavors.  The off-diagonal parts,
however, are much more tightly constrained by the decay process $\nu_i
\to \nu_j \phi$ \cite{Hannestad:2005ex}.

At $z \sim 1500$ the universe is matter dominated and to a good
approximation $H \propto T^{3/2}$. Since for scattering the rate is
$\Gamma \sim g^4 T$, we can translate the condition for strong
interaction, $\Gamma/H \gtrsim 1$, to a bound on $g$
\cite{Hannestad:2005ex}. Since $\Gamma/H \propto g^4 T^{-1/2}$, and
in the previous paper we effectively used $z_i = 1088$ to obtain $g <
10^{-7}$ we now get $g < 10^{-7} (1500/1088)^{1/8} \sim 1.05 \times
10^{-7}$, i.e.\ a negligible 5\% difference compared with our
previous result.

It should be noted that for masses below the recombination
temperature, $m \lesssim T_R \sim 0.3$ eV, our bound applies equally
well to the case of neutrino decay and inverse
decay~\cite{Hannestad:2005ex}. In fact, we can now make the bound more
quantitative by adding that at 95\% C.L. the interaction cannot be
very strong before $z_i \sim 1500$. Roughly this translates to a bound
on the lifetime (again scaling from our previous limit derived using
$z_i = 1088$) of
\begin{equation}
\tau > 1.0 \times 10^{10} \, {\rm s} \, \left(\frac{m}{50 \, {\rm
meV}}\right)^3.
\end{equation}
This limit is slightly weaker than before~\cite{Hannestad:2005ex}
because of the slightly more conservative assumption about $z_i$, but
it remains by far the most restrictive bound on invisible decays of
low-mass neutrinos.

\section{Discussion}                            \label{sec:discussion}

We have updated bounds on the coupling strength between neutrinos and
a new, light pseudo-scalar, $\phi$, using the latest cosmological
data. Performing a slightly more refined calculation than in an
earlier paper by two of us~\cite{Hannestad:2005ex} we find
essentially unchanged constraints.

One way to improve this limit in future is by actually detecting
neutrino hot dark matter in cosmological precision data.
In decay scenarios involving massless pseudoscalars and for a mass of
50~meV, the lifetime limit would improve by some six orders of
magnitude~\cite{serpico}.

Our more general conclusion is that neutrinos which are strongly
interacting around recombination are strongly disfavored by data. The
present data strongly support the conclusion that the cosmic neutrino
background (i)~exists around the epoch of recombination and (ii)~its
fluctuations do have an anisotropic stress component. In future CMB
data alone will likely suffice to reach the same or better sensitivity
so that the bound on $g$ can be expected to improve significantly
\cite{Friedland:2007vv, DeBernardis:2008ys}.

\vspace*{-0.5cm}
\section*{Acknowledgments}
\vspace*{-0.2cm}

We acknowledge use of computing resources from the Danish Center for
Scientific Computing (DCSC). GGR acknowledges partial support by the
Deutsche Forschungsgemeinschaft (grant TR-27) and by the Cluster of
Excellence ``Origin and Structure of the Universe.''


\end{document}